\documentclass[a4paper,10pt,twoside]{cpc-hepnp}
\usepackage{CJKutf8,upgreek,fancyhdr}
\usepackage{multicol}
\usepackage{graphicx}
\usepackage{booktabs}
\usepackage{amssymb,bm,mathrsfs,bbm,amscd}
\usepackage[tbtags]{amsmath}
\usepackage{lastpage}

\newcommand{\be}{\begin{equation}}
\newcommand{\ee}{\end{equation}}
\newcommand{\bea}{\begin{eqnarray}}
\newcommand{\eea}{\end{eqnarray}}

\def\met{\slash{\!\!\!\!\!\!E}_{\text{T}}}

\def\to{\rightarrow}

\def\to{\rightarrow}

\newcommand{\lsim}{\mathrel{\mathop{\kern 0pt \rlap
  {\raise.2ex\hbox{$<$}}}
  \lower.9ex\hbox{\kern-.190em $\sim$}}}
\newcommand{\gsim}{\mathrel{\mathop{\kern 0pt \rlap
  {\raise.2ex\hbox{$>$}}}
  \lower.9ex\hbox{\kern-.190em $\sim$}}}

\begin{document}
\begin{CJK*}{UTF8}{gbsn}

\fancyhead[c]{\small Chinese Physics C~~~Vol. xx, No. x (2020) xxxxxx}
\fancyfoot[C]{\small 010201-\thepage}

\footnotetext[0]{Received dd mm 2020}

\title{Investigating Bottom-Quark Yukawa Interaction at Higgs Factory
\thanks{The work of JG is sponsored by the National Natural Science Foundation of China under the Grant No. 11875189 and No. 11835005, and by the MOE Key Lab for Particle Physics, Astrophysics and Cosmology. The research of QB, KC, YL and HZ is supported by the funding from the Institute of High Energy Physics, Chinese Academy of Sciences with Contract No. Y6515580U1, and the funding from Chinese Academy of Sciences with Contract No. Y8291120K2. JG and HZ are pleased to recognize the support and the hospitality of the Center for High Energy Physics at Peking University. }}

\author{%
Qi Bi (毕琪)$^{1,2},$\email{biqi@ihep.ac.cn}%
\quad      Kangyu Chai (柴康钰)$^{1,2},$\email{chaikangyu@ihep.ac.cn}%
\quad      Jun Gao (高俊)$^{3,4},$\email{jung49@sjtu.edu.cn}%
\quad      Yiming Liu (刘一鸣)$^{1,2},$\email{liuym@ihep.ac.cn}%
\quad Hao Zhang (张昊)$^{1,2,4},$\email{zhanghao@ihep.ac.cn}%
}
\maketitle

\address{%
$^1$ Institute of High Energy Physics, Chinese Academy of Sciences, Beijing 100049, China\\
$^2$ School of Physics, University of Chinese Academy of Science, Beijing 100049, China\\
$^3$ INPAC, Shanghai Key Laboratory for Particle Physics and Cosmology, School of Physics and Astronomy, Shanghai Jiao-Tong University, Shanghai 200240, China
\\
$^4$ Center for High Energy Physics, Peking University, Beijing 100871, China
}

\begin{abstract}
Measuring the fermion Yukawa coupling constants is important for understanding the origin of the fermion masses and its relationship to the spontaneously electroweak symmetry breaking. On the other hand, some new physics models will change the Lorentz structure of the Yukawa interactions between the standard model (SM) fermions and the SM-like Higgs boson even in their decoupling limit. Thus the precisely measurement of the fermion Yukawa interactions is a powerful tool of new physics searching in the decoupling limit. In this work, we show the possibility of investigating the Lorentz structure of the bottom-quark Yukawa interaction with the 125GeV SM-like Higgs boson at future $e^+e^-$ colliders.
\end{abstract}

\begin{keyword}
Higgs, bottom-quark Yukawa interaction, BSM, Higgs factory
\end{keyword}

\begin{pacs}
12.60.Fr, 13.66.Fg, 14.80.Bn
\end{pacs}

\footnotetext[0]{\hspace*{-3mm}\raisebox{0.3ex}{$\scriptstyle\copyright$}2013 Chinese Physical Society and the Institute of High Energy Physics of the Chinese Academy of Sciences and the Institute of Modern Physics of the Chinese Academy of Sciences and IOP Publishing Ltd}%

\begin{multicols}{2}

\section{Introduction}
\label{sec-intro}

After the discovery of the SM-like Higgs boson at the Large Hadron Collider (LHC) \cite{Aad:2012tfa,Chatrchyan:2012xdj}, particle physicists pay more and more attention on the investigation of the properties of the SM-like Higgs boson. With the theoretical and experimental uncertainties, most of the results are in consistent with the SM predictions \cite{Sirunyan:2018koj,Aad:2019mbh}.

To verifying the prediction of the SM, it is not enough to check the strength of interactions between the SM-like Higgs boson and other SM particles. People need to investigate the Lorentz structure and the coupling constants associated with each possible Lorentz structure. For example, the generic form of the interaction between the SM-like Higgs boson and the SM fermions is 
\bea
&&\mathscr{L}_{Y_f}=-y_fh\bar \psi_f(\cos\alpha_f+i\gamma_5\sin\alpha_f)\psi_f,
\label{eq:1}\\
&&y_f>0,~\alpha_f\in (-\pi,\pi],~f=e,\mu,\tau,u,d,c,s,t,b.\nonumber
\eea
In the SM, we have $y_f=y_f^{\text{SM}}=m_f/(\sqrt2v)$ ($v=174$GeV is the vacuum expectation of the SM Higgs field) and $\alpha_f=0$ for massive SM fermions. Although the phase angle $\alpha_f$ could be removed by a redefinition of the fermion field
\be
\psi_f\to \psi_f^\prime=e^{-i\alpha_f\gamma_5/2}\psi_f
\label{eq:2}
\ee
for massless fermions, such redefinition will not work for massive fermions since their phases have been fixed by the mass $m_f\in\mathbb R^+$ in the Lagrangian $\bar\psi_f(i\slash{\!\!\!\!\!\!D}-m_f)\psi_f$ of free fermion fields. Thus, either $y_f\neq m_f/(\sqrt2v)$ or $\alpha_f\neq0$ will be the evidence of the new physics (NP) beyond the SM.

Due to the large $y_t$, the measurement of the phase angle in the top-Higgs interaction $\alpha_t$ is relatively easy and proposed in a lot of works (for example, see \cite{Gunion:1996xu,Biswas:2012bd,He:2014xla,Boudjema:2015nda,Li:2015kxc,Khachatryan:2015ota,Mileo:2016mxg,AmorDosSantos:2017ayi,Gouveia:2018xfp,Vryonidou:2018eyv,Boselli:2018zxr,Durieux:2018tev,Durieux:2018ggn,Ma:2018ott,Sirunyan:2018lzm,Ren:2019xhp}). However, the $\alpha_f$'s of the down-type fermions are also very interesting and important from the theoretical point of view. A well known example is the ``wrong-sign limit'' in some kinds of the two-Higgs-doublet model (2HDM). Without any other deviation from the predictions of the SM, $\alpha_b\approx\pi$ (because $y_b$ is the largest $y_f$ in the down-type fermions, $\alpha_b$ is probably the easiest one to be measured) will be strong hint of these kinds of NP models.

Many efforts have been done for measuring $\alpha_b$. Although the direct measurement is very challenge at the LHC \cite{Aaboud:2018txb,ATL-PHYS-PUB-2015-043}, it could be measured in the electric dipole moments (EDM) experiments indirectly \cite{Brod:2013cka,Chien:2015xha,Brod:2018pli}, or at the LHC with additional model-dependent assumptions (e.g., in the frame of 2HDM \cite{Ferreira:2014naa,Fontes:2014tga,Ferreira:2014dya,Biswas:2015zgk,Modak:2016cdm,Han:2017etg,Coyle:2018ydo,Modak:2018zro,Chen:2018shg,Chen:2019rdk}). The constraint from the indirect measurement is strong but suffered by the potentially contributions from exotic degree of freedoms in NP. For this reason, a direct, model-independent measurement is still necessary.

In this work, we investigate the possibility of measuring $\alpha_b$ directly and model-independently at future Higgs factory.

\section{The phenomenology of the bottom-quark Yukawa interaction}
To the leading order, the effective Lagrangian in the Eq. (\ref{eq:1}) modifies the $h\to b\bar b$ decay width to
\be
\Gamma(h\to b\bar b)=\Gamma(h\to b\bar b)^{\text{SM}}\left(\frac{y_b}{y_b^{\text{SM}}}\right)^2\left(\cos^2\alpha_b+\beta_b^{-2}\sin^2\alpha_b\right),
\label{eq:3}
\ee
where $\beta_b\equiv \sqrt{1-4m_b^2/m_h^2}$. The precisely measurement of the decay branching ratio can only constrain the combination 
\bea
&&\left(\frac{y_b}{y_b^{\text{SM}}}\right)^2\left(\cos^2\alpha_b+\beta_b^{-2}\sin^2\alpha_b\right)\nonumber\\
\sim&&\left(\frac{y_b}{y_b^{\text{SM}}}\right)^2\left(1+\frac{4m_b^2}{m_h^2}\sin^2\alpha_b\right)\nonumber\\
=&&\left(\frac{y_b^{\text{SM}}+\delta y_b}{y_b^{\text{SM}}}\right)^2\left(1+0.0058\sin^2\alpha_b\right)\nonumber\\
\sim&&1+2\left(\frac{\delta y_b}{y_b^{\text{SM}}}\right)+\left(\frac{\delta y_b}{y_b^{\text{SM}}}\right)^2+0.0058\sin^2\alpha_b
\label{eq:4}
\eea
of the $y_b$ and $\alpha_b$, in which the contribution from $\alpha_b$ is numerically small. Even if we keep $y_b=y_b^{\text{SM}}$, the partial width will be in the region of $\Gamma(h\to b\bar b)^{\text{SM}}(1.0029\pm0.29\%)$. This small discrepancy is just below the sensitivity at the Higgs factories \cite{CEPCStudyGroup:2018ghi,An:2018dwb,Abada:2019zxq}. So we have to look for other kinematic variables which are sensitive to $\alpha_b$.

To measure $\alpha_b$, we consider the interference effect in the $h\to \bar bbg$ process, whose Feynman diagrams are shown in Fig. \ref{fig:1}. 
\begin{center}
\includegraphics[width=8.8cm]{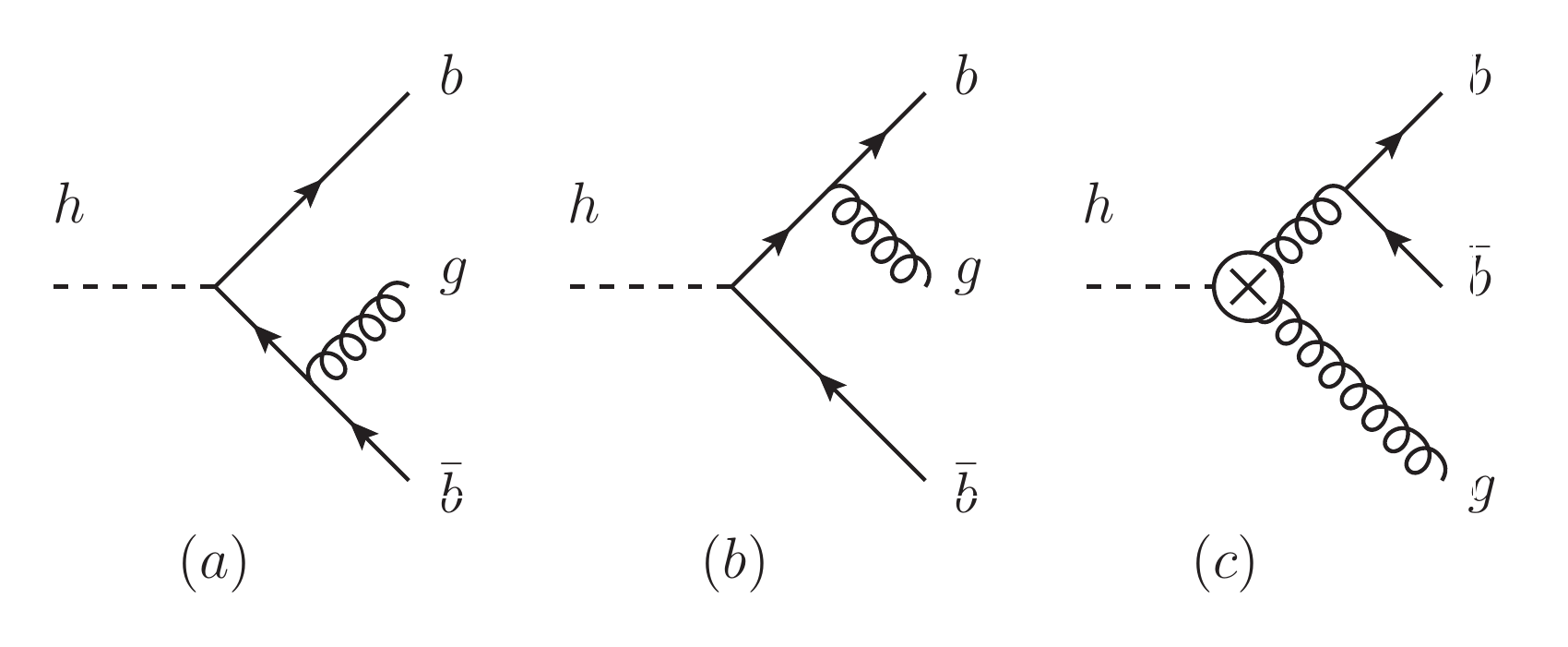}
\figcaption{\label{fig:1}The Feynman diagrams which are used to measure the relative sign between the bottom-quark Yukawa coupling constant and the weak interaction gauge coupling constant. }  
\end{center}
The transition amplitude can be written as
\be
\mathscr M=e^{\pm i\alpha_b}\mathscr M_1+\mathscr M_2
\label{eq:5}
\ee
where $\mathscr M_1$ represents the contribution from the Feynman diagrams (a) and (b), $\mathscr M_2$ represents the contribution from the Feynman diagram (c), and they are both $\alpha_b$-independent. In the Eq. (\ref{eq:5}), the sign before the phase angle $\alpha_b$ depends on the chirality configuration of the $b\bar b$ in the final state.

Because the $hb\bar b$ vertex flips the chirality of the fermion line, while the $gb\bar b$ does not, if the $b$-quark is massless, the interference term will vanish. It can only appear when $b$-quark is massive in which case the chirality is not a good quantum number. The $\mathscr M_1$ and $\mathscr M_2$ can be  non-zero at the same time due to the mass insertion effect. This technical analysis can be understood easily. Since in the massless limit the chiral symmetry restores and one can remove the $\alpha_b$ with the symmetry transformation Eq. (\ref{eq:2}), $\alpha_b$ should not have any observable effect in this limit. So any observable effect of $\alpha_b$ is expected to be proportional to $m_b$.

Our next mission is finding the phase space region where the interference effect is large. It will guide us to design a suitable observable and cuts. The relative size of the interference effect can be described by the ratio between the interference term and the non-interference terms
\bea
&&\frac{e^{\pm i\alpha_b}\mathscr M_1\mathscr M_2^*+e^{\mp i\alpha_b}\mathscr M_1^*\mathscr M_2}{|\mathscr M_1|^2+|\mathscr M_2|^2}\nonumber\\
&=&2\cos(\pm\alpha_b+\phi)\frac{|\mathscr M_1|\cdot|\mathscr M_2|}{|\mathscr M_1|^2+|\mathscr M_2|^2},
\label{eq:6}
\eea
where $\phi$ is phase angle of $\mathscr M_1\mathscr M_2^*$. As a matter of fact, we can only measure $\alpha_b+\phi$ with this process. However, the effective $hgg$ vertex 
\be
\left(\frac{\alpha_s}{12\sqrt 2\pi v}+\frac{c_{hgg}}{\Lambda}\right)hG_{\mu\nu}^aG^{a,\mu\nu}+\frac{\tilde c_{hgg}}{\Lambda}hG_{\mu\nu}^a\tilde G^{a,\mu\nu}
\label{eq:7}
\ee
can be independently, precisely measured at the LHC \cite{Dolan:2014upa,Kobakhidze:2016mfx,Bernlochner:2018opw,Englert:2019xhk,Kraus:2019myc}, so the model dependence from this part is little. This is another advantage of this process. In our work, we choose the SM value, $c_{hgg}=\tilde c_{hgg}=0$ in the low energy limit. To get a significant modulation effect, we need to find the phase space region where $|\mathscr M_1|\cdot|\mathscr M_2|/(|\mathscr M_1|^2+|\mathscr M_2|^2)$ is large. It is obviously that this quantity reaches its maximal value when $|\mathscr M_1|=|\mathscr M_2|$. Because $y_b>\alpha_sm_h/(12\sqrt 2\pi v)$, we have generically $|\mathscr M_1|>|\mathscr M_2|$. So we should focus on the phase space region where the $\mathscr M_2$ is more enhanced. Certainly, it is the the region where the $b\bar b$ is collinear. Because the $\mathscr M_2$ obtains a large QCD collinear divergence in this region and is largely enhanced, while the $\mathscr M_1$ obtains no QCD divergence there. Guiding by this analysis, we define an observable as
\be
\zeta_{H}\equiv\frac{2E_{b_1}E_{b_2}}{E_{b_1}^2+E_{b_1}^2}\cos\theta_{b_1b_2},
\label{eq:8}
\ee
where $E_{b_i}$ is the energy of the $i$th $b$-jet in the Higgs rest-frame, $\theta_{b_1b_2}$ is the open angle between the 2 $b$-jets in the Higgs-rest frame.

A straightforward calculation gives the differential partial decay width (to the order of $m_b$)\footnote{We would like to emphasize that the $m_b$ in the formula, as the mass of the bottom-quark, only comes from the propagator of the bottom-quark, while the $y_b$ is from the interaction vertices. We do not use the relation $y_b=m_b/(\sqrt2 v)$ for two reasons. First, it is a relation in the SM which might be broken in NP models. Second, even in the SM, this relation is not good enough when people want to mimic some higher order effect. }
\bea
\frac{d^2\Gamma}{dx_{13}dx_{23}}&=&\frac{y_b^2m_h\alpha_s}{4\pi^2}\biggl\{\Pi_{11}(x_{13},x_{23})+2\Pi_{12}(x_{13},x_{23})\nonumber\\
&&\times\frac{m_b}{m_h}r\cos\alpha_b+\Pi_{22}(x_{13},x_{23})r^2\biggr\},\label{eq:9}\\
\Pi_{11}(x_{13},x_{23})&=&\frac{1+(1-x_{13}-x_{23})^2}{x_{13}x_{23}},\label{eq:10}\\
\Pi_{12}(x_{13},x_{23})&=&\frac{(x_{13}+x_{23})(x_{13}-x_{23})^2+4x_{13}x_{23}}{x_{13}x_{23}(1-x_{13}-x_{23})},\label{eq:11}\\
\Pi_{22}(x_{13},x_{23})&=&\frac{x_{13}^2+x_{23}^2}{(1-x_{13}-x_{23})},\label{eq:12}
\eea
where 
\be
r\equiv \frac{\alpha_s}{6\sqrt2\pi y_b}\left(\frac{m_h}{v}\right)\sim\frac{1}{4},
\label{eq:13}
\ee
$x_{13}=(p_b+p_g)^2/m_h^2$, $x_{23}=(p_{\bar b}+p_g)^2/m_h^2$, in which $p_b,p_{\bar b}$ and $p_g$ is the four momentum of the bottom-quark, anti-bottom-quark and gluon in the Higgs-rest frame, respectively. In this formula, the $\Pi_{ij}$ term is from the amplitude square $(\mathscr M_i^*\mathscr M_j+\mathscr M_i\mathscr M_j^*)/(1+\delta_{ij})$ term. It is easy to verify our intuitive analysis with this formula.

\section{The collider phenomenology}
In this section, we investigate the collider phenomenology at future Higgs factory \cite{CEPCStudyGroup:2018ghi,Abada:2019zxq}. The lepton collider is designed to run with 240GeV collision energy with roughly 5fb$^{-1}$ integrated luminosity\footnote{In our simulation, we set the integrated luminosity for 240GeV Higgs factory to be 5.6fb$^{-1}$ following \cite{An:2018dwb}. The result for 5fb$^{-1}$ integrated luminosity 240GeV Higgs factory will be very closed to the result given in this work, and easy to get by a simple rescaling. }. Some of them also have plan to run 365GeV collision energy with roughly 1.5fb$^{-1}$ integrated luminosity \cite{Abada:2019zxq}. We will give the results of parton level collider simulation for both 240GeV and 365GeV lepton collider here.
 
\subsection{The 240GeV Higgs Factory}
We generate parton level signal and background events at 240GeV $e^+e^-$ collider using MadGraph$\_$aMC@NLO \cite{Alwall:2014hca} with the initial state radiation (ISR) effects \cite{Chen:2017gzv}. To include the NNLO corrections to the cross section, the total cross section of $e^+e^-\to Zh$ is rescaled to the suggested value in \cite{Gong:2016jys,Sun:2016bel,Chen:2018xau}. We analyze both leptonic and hadronic decay modes of the $Z$ boson. The interference effect between the Higgs strahlung process and the $Z$-boson fusion process in the $e^+e^-$ decay case of $Z$ boson is considered in our analysis. The jet algorithm is the $ee\_$kt (Durham) algorithm in which the distance between the object $i$ and $j$ is defined as \cite{Catani:1991hj}
\begin{equation}
d_{ij}\equiv2\left(1-\cos\theta_{ij}\right)\frac{\min\left(E_i^2,E_j^2\right)}{s},
\label{eq:13}
\end{equation}
where $s$ is the square of the center-of-mass frame energy, $E_{i}$ is the energy of the $i$th jet, $\theta_{ij}$ is the angle  opened by the $i$th and $j$th jet.

We add pre-selection cuts when we generate the parton level event
\bea
&&|\eta_{jet,\ell^\pm}|<2.3,~\Delta R_{ij}>0.1, \Delta R_{i\ell}>0.2,\nonumber\\
&&E_{jet}>10{\text{GeV}},~E_{\ell^\pm}>5{\text{GeV}}.\nonumber
\eea
The parameters of the smearing effects for different particles are chosen to be \cite{CEPCStudyGroup:2018ghi}
\bea
\frac{\sigma(E_{jet})}{E_{jet}}&=&\frac{0.60}{\sqrt{E_{jet}/{\text{GeV}}}}\oplus 0.01,\nonumber\\
\frac{\sigma(E_{e^\pm,\gamma})}{E_{e^\pm,\gamma}}&=&\frac{0.16}{\sqrt{E_{e^\pm,\gamma}/{\text{GeV}}}}\oplus 0.01,\nonumber\\
\sigma\left(\frac{1}{p_{{\text T},\mu^\pm}}\right)&=&2\times10^{-5}~{\text{GeV}}^{-1}\oplus\frac{0.001}{p_{\mu^\pm}\sin^{3/2}\theta_{\mu^\pm}},\nonumber
\eea

\subsubsection{Leptonic Decaying $Z$}
After adding the smearing effects, we require the objects satisfy\footnote{The value of the $d_{ij}$ cut is based on the assumption that the future lepton collider has a resolution at least as good as the LEP \cite{Barate:1998gx,Abbiendi:1998rd}. }
\bea
&&|\cos\theta_{{jet},\ell^\pm}|<0.98,~d_{ij}>0.002, E_{jet}>15{\text{GeV}}, \nonumber\\
&&\Delta R_{i\ell^\pm}>0.2,~E_{\ell^\pm}>10{\text{GeV}}.
\nonumber
\eea
The $b$-tagging efficiency is chosen to be 80\%, while the mis-tagging rate from charm jet (light jet) is 10\% (1\%). After the preselection cuts, we require the signal events contain exact 2 $b$-tagged jets, 1 non-$b$ jet, a pair of opposite sign same flavor charged leptons, and 
\bea
&&|m_{\mu^+\mu^-}-m_Z|<10{\text{GeV}},~~|m_{e^+e^-}-m_Z|<15{\text{GeV}},\nonumber\\
&&\theta_{\ell^+\ell^-}>80^\circ,~~~~\met<10{\text{GeV}},\nonumber\\
&&124.5{\text{GeV}}<m_{\text{r}}(\mu^+\mu^-)<130{\text{GeV}},~{\text{for}}~\mu^+\mu^-~{\text{channel}},\nonumber\\
&&118~{\text{GeV}}<m_{\text{r}}(e^+e^-)<140{\text{GeV}},~{\text{for}}~e^+e^-~{\text{channel}},\nonumber
\eea
where the recoil mass $m_{\text{r}}(ij)$ is defined as
\be
m_{\text{r}}(ij)\equiv \sqrt{ s-2\sqrt s (E_i+E_j)+(p_i+p_j)^2}.
\label{eq:14}
\ee

The dominant SM background processes for $Z\to\ell^+\ell^-$ channel is
\bea
e^+e^-&\to& \ell^+\ell^-b\bar b j\nonumber\\
e^+e^-&\to& \ell^+\ell^-c\bar c j\nonumber\\
e^+e^-&\to& \ell^+\ell^-j j j\nonumber\\
e^+e^-&\to& \ell^+\ell^- h(\to c\bar c j)\nonumber\\
e^+e^-&\to& \ell^+\ell^- h(\to j j j)\nonumber
\eea
The kinematic cut on the recoil mass of $\ell^+\ell^-$ can remove most of the background events from the first three SM processes, while the last two can pass this cut. However, the last two background will be suppressed by the charm-jet and light jet mistagging rate. 

In our analysis, the 4-momentum of the Higgs boson is reconstruct by summing the 4-momentum of the three jets from the Higgs boson decay, but not the recoil momentum of the dilepton system. When the 2 $b$-jets from the Higgs boson decay are nearly collinear and the $b\bar b$-system and the gluon jet from the Higgs boson decay is nearly back-to-back, $\zeta_H$ goes to its maximum value, +1. In Fig. \ref{fig:2}, we show the $\zeta_H$ distributions for the SM backgrounds and the signal with different values of $\alpha_b$. The behavior of the distribution, especially in the last several bins, is in consistent with our intuitive analysis.
\begin{center}
\includegraphics[width=7.5cm]{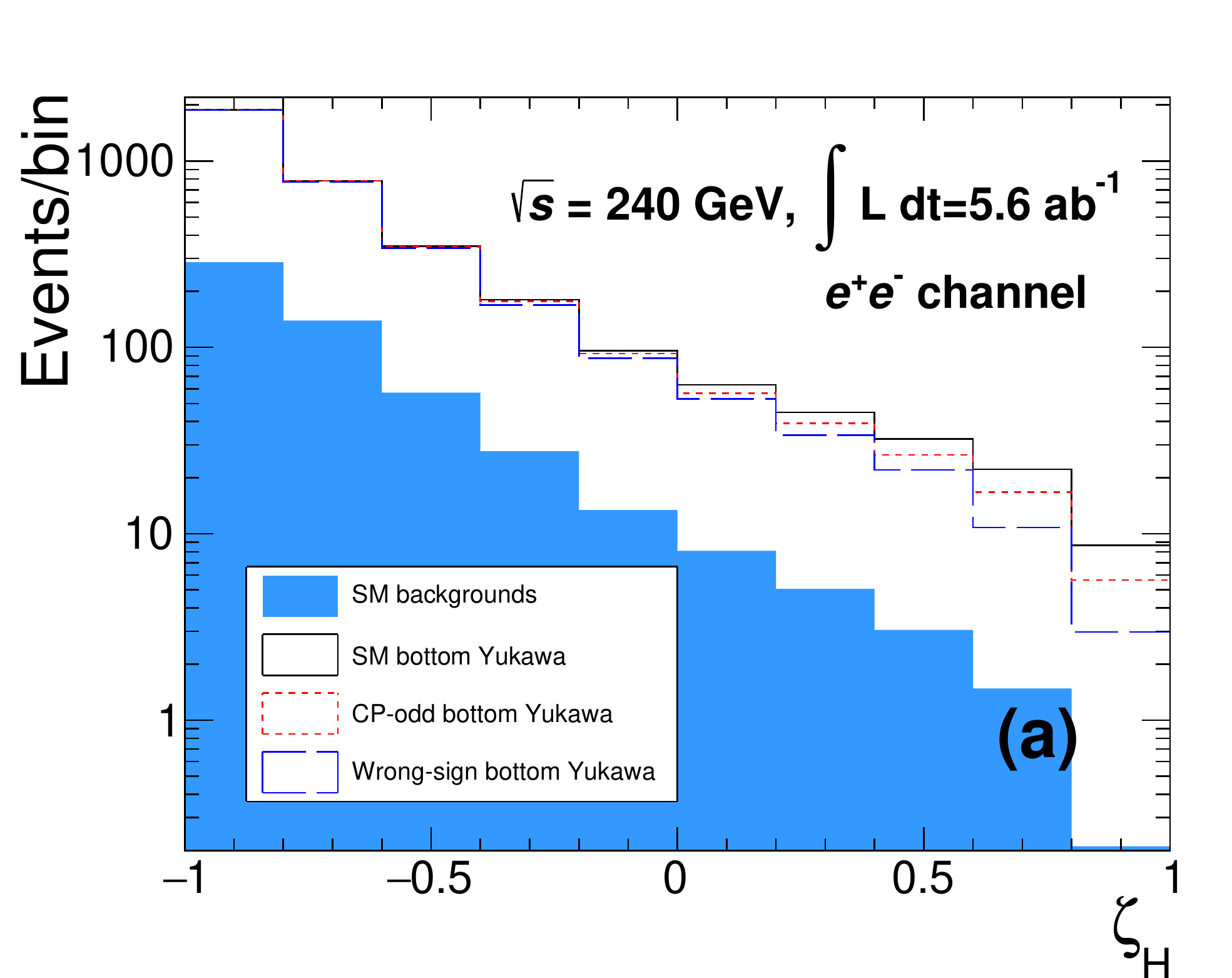}\\
\includegraphics[width=7.5cm]{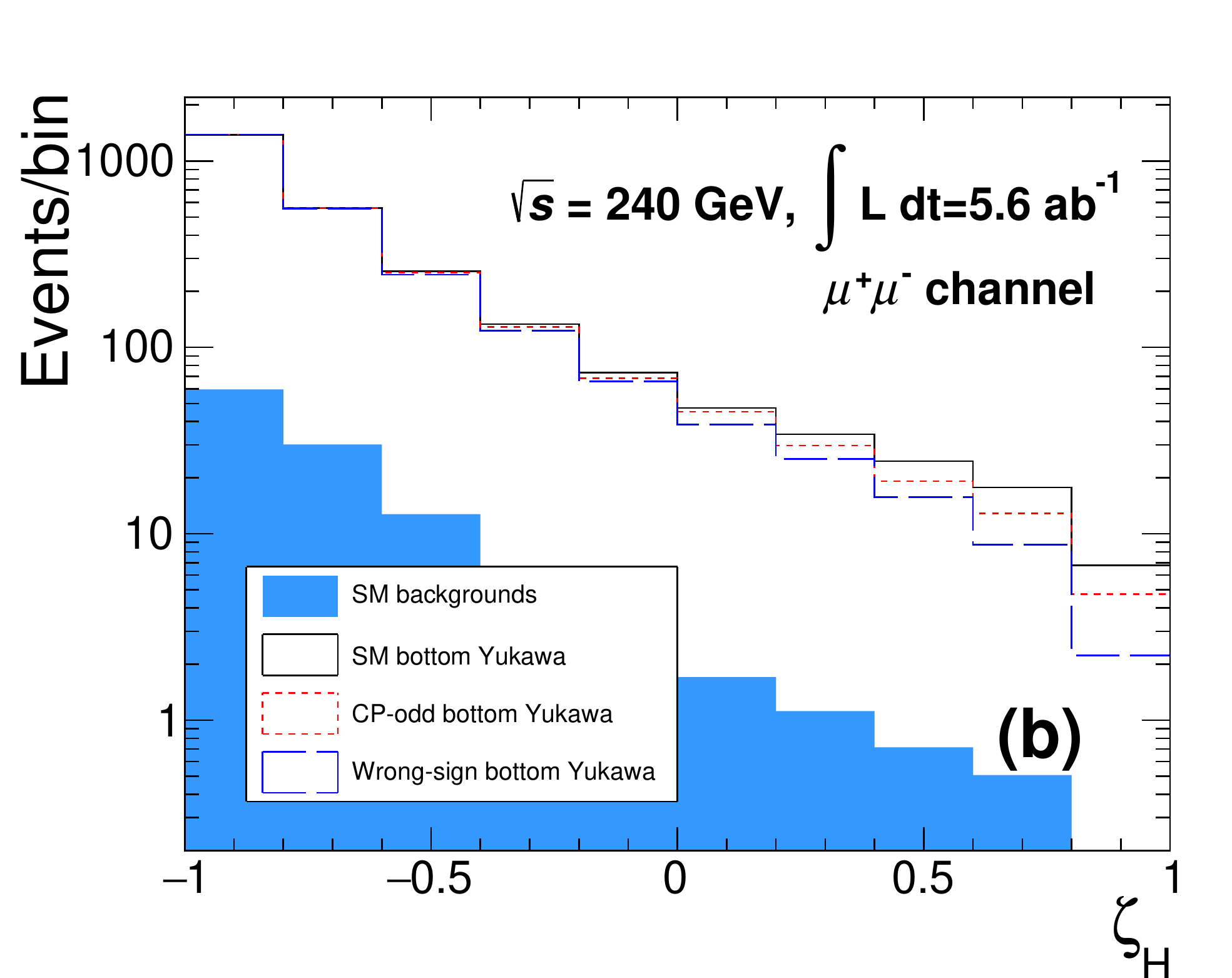}\\
\includegraphics[width=7.5cm]{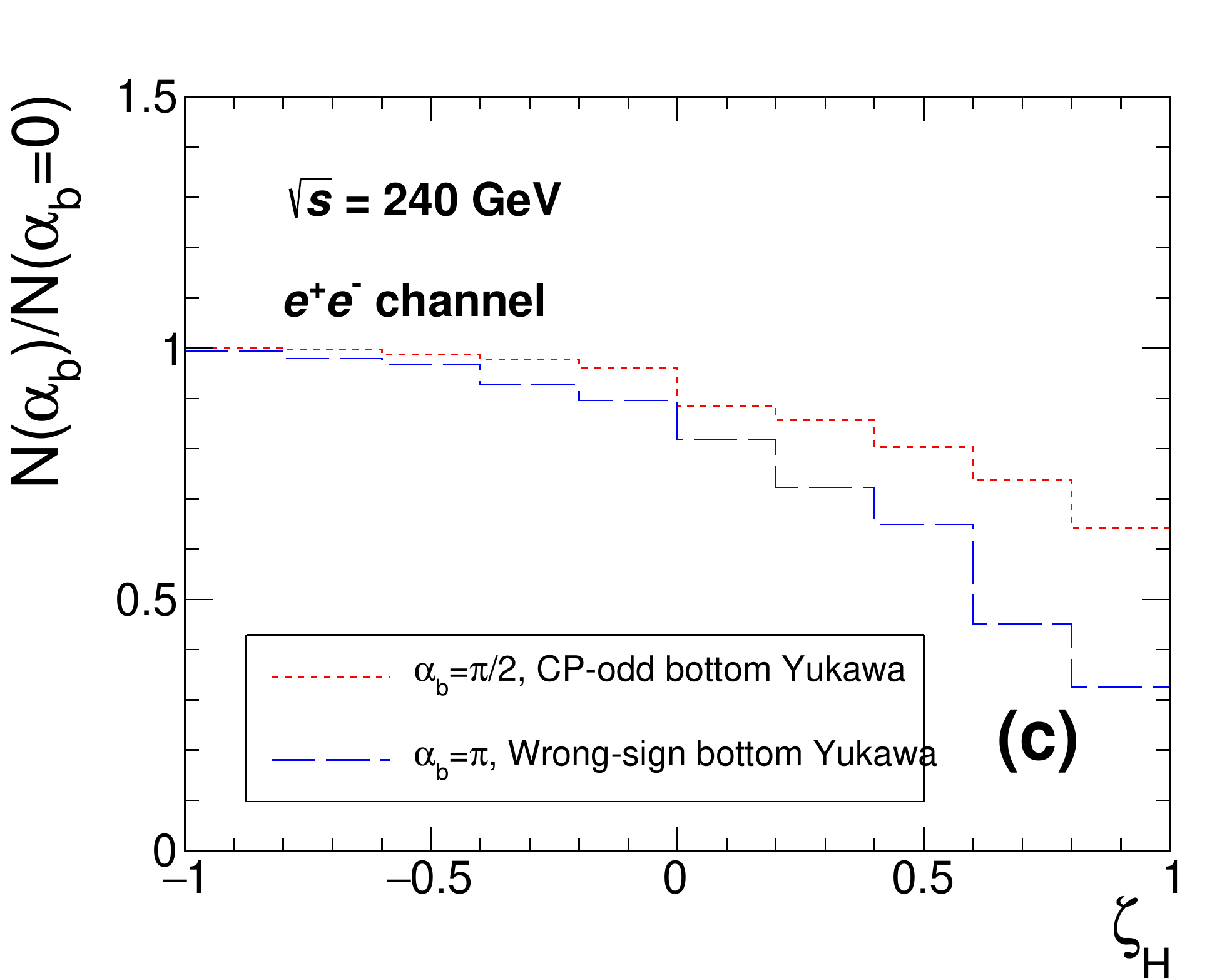}\\
\includegraphics[width=7.5cm]{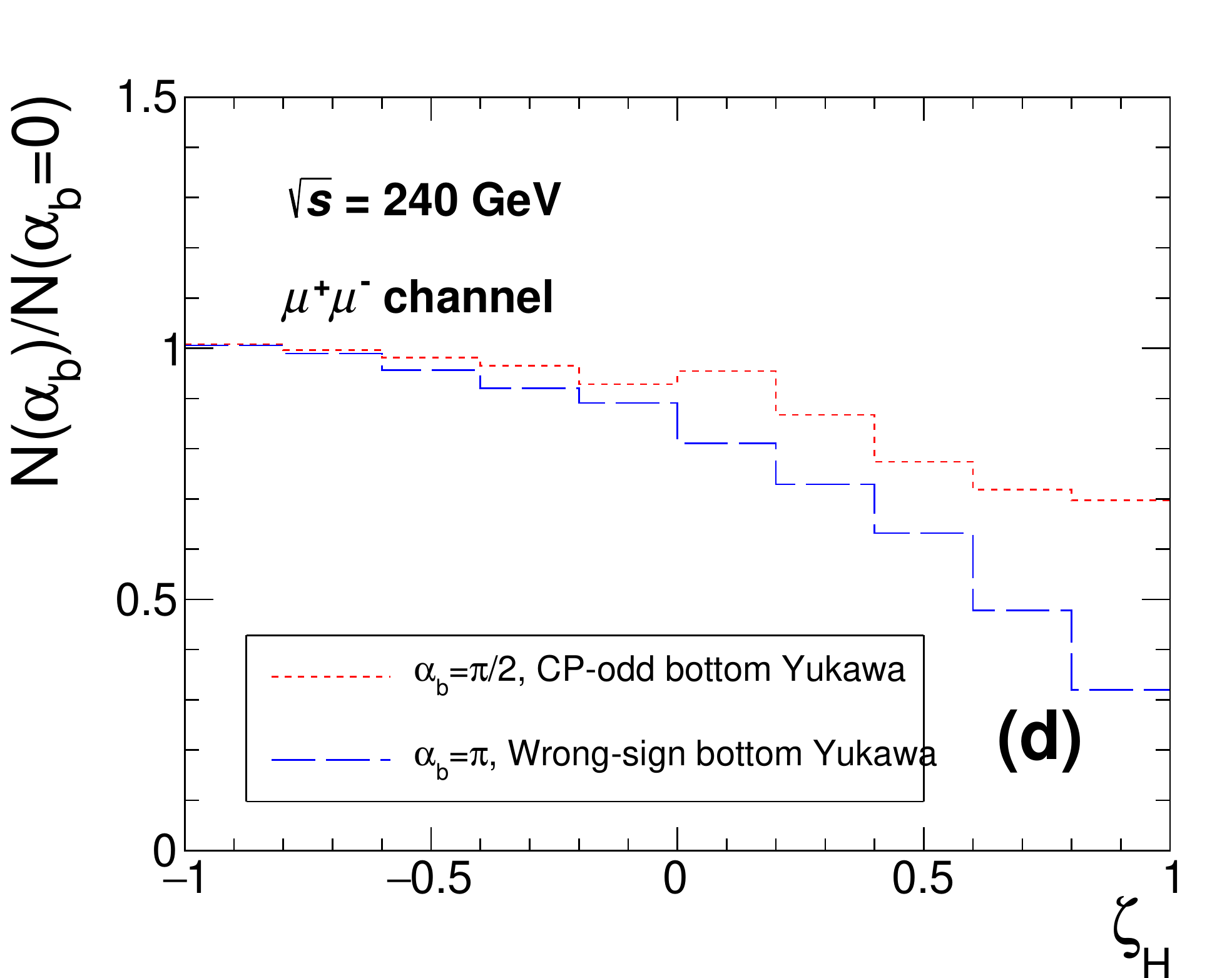}
\figcaption{\label{fig:2}The $\zeta_H$ distributions for the SM background, the SM bottom-quark Yukawa interaction ($\alpha_b=0$), bottom-quark Yukawa interaction with CP-odd scalar ($\alpha_b=\pi/2$), and the wrong-sign bottom-quark Yukawa interaction ($\alpha_b=\pi$) at 240 GeV Higgs factory with 5.6 ab$^{-1}$ integrated luminosity. (a) The $\zeta_H$ distribution of $Z\to e^+e^-$ channel; (b) The $\zeta_H$ distribution of $Z\to \mu^+\mu^-$ channel; (c) The ratio of the event rates with respect to the SM case ($\alpha_b=0$) of $Z\to e^+e^-$ channel; (d) The ratio of the event rates with respect to the SM case ($\alpha_b=0$) of $Z\to \mu^+\mu^-$ channel.}  
\end{center}

\subsubsection{Hadronic Decaying $Z$}
Although the analysis is more complicate than the channels in which the $Z$ boson decays leptonically, the branching ratio of the hadronic decay mode of $Z$ boson is much larger. Thus it is worth to make effort to include the information from this channel. After adding the smearing effects, we require the objects satisfy
\be
|\cos\theta_i|<0.98,~d_{ij}>0.002, E_{jet}>15{\text{GeV}}, \met<10{\text{GeV}}.\nonumber
\ee
To avoid a too aggressive estimation in the jet-rich environment, for this mode, we assume that the $b$-tagging efficiency is 60\% (lower than the leptonic channel), while the mis-tagging rate from charm jet (light jet) is 10\% (1\%). After the preselection cuts, we require the signal events contain at least 2 $b$-tagged jets, and 5 jets in total. To reconstruct the Higgs boson and the $Z$ boson, we use the likelihood method. The distribution of the truth reconstructed $Z$-boson mass, Higgs boson mass, $Z$-boson recoil mass and Higgs boson recoil mass are 
\bea
L_Z(m)&=&P(m;91.0{\text{GeV}},6.19{\text{GeV}}),\label{eq:16}\\
L_h(m)&=&P(m;125.3{\text{GeV}},6.54{\text{GeV}}),\label{eq:17}\\
L_{rZ}(m)&=&P(m;126.7{\text{GeV}},8.43{\text{GeV}}),\label{eq:18}\\
L_{rh}(m)&=&P(m;93.0{\text{GeV}},10.56{\text{GeV}}),\label{eq:19}
\eea
respectively, where
\be
P(x;\mu,\sigma)=\frac{1}{\sqrt{2\pi}\sigma}\exp\left[-\frac{(x-\mu)^2}{2\sigma^2}\right]
\label{eq:20}
\ee
is the standard probability distribution function (p.d.f) of the normal distribution. We minimize a discriminator defined as 
\bea
\Delta&=&-2\ln L_Z(m_{i_1i_2})-2\ln L_h(m_{i_3i_4i_5})\nonumber\\
&&-2\ln L_{rZ}(m_{{\text{recoil}}}(i_1i_2))-2\ln L_{rh}(m_{\text{recoil}}(i_3i_4i_5))\nonumber\\
&&-70B({i_3})-70B({i_4})+100B({i_5}),
\label{eq:21}
\eea
where $i_1,\cdots,i_5$ is a permutation of the 5 jets, $m_{i\cdots j}$ is the invariant mass of the $i$th, $\cdots$, and the $j$th jet,  $m_{\text{recoil}}(i\cdots j)$ is the recoil mass of the $i$th, $\cdots$, and the $j$th jet, $B(i)$ is 1 (0) if the $i$th jet is tagged (not) to be a $b$-jet. If $i_1,\cdots,i_5$ gives the minimum $\Delta$, we treat $j_{i_1},j_{i_2}$ as jets from $Z$ decay, $j_{i_3},j_{i_4}$ as the $b$-jets from the Higgs boson decay, and $j_{i_5}$ as the gluon from the Higgs boson decay. For the signal events, the reconstruction efficiency is $\sim80\%$. We require there is at least 2 $b$-jets in $j_{i_3},j_{i_4}$ and $j_{i_5}$, $\Delta<45$ and $120^\circ<\theta_{i_1i_2}<150^\circ$. 

The dominant SM background processes for $Z\to jj$ channel is
\bea
e^+e^-&\to& j j j j j\nonumber\\
e^+e^-&\to& j j h(\to c\bar c j)\nonumber\\
e^+e^-&\to& j j h(\to j j j)\nonumber
\eea
After the reconstruction, we can get the $\zeta_H$ distribution which is shown in Fig. \ref{fig:3}, we show the $\zeta_H$ distributions for the residue SM backgrounds and the signal with different values of $\alpha_b$. 
\begin{center}
\includegraphics[width=7.5cm]{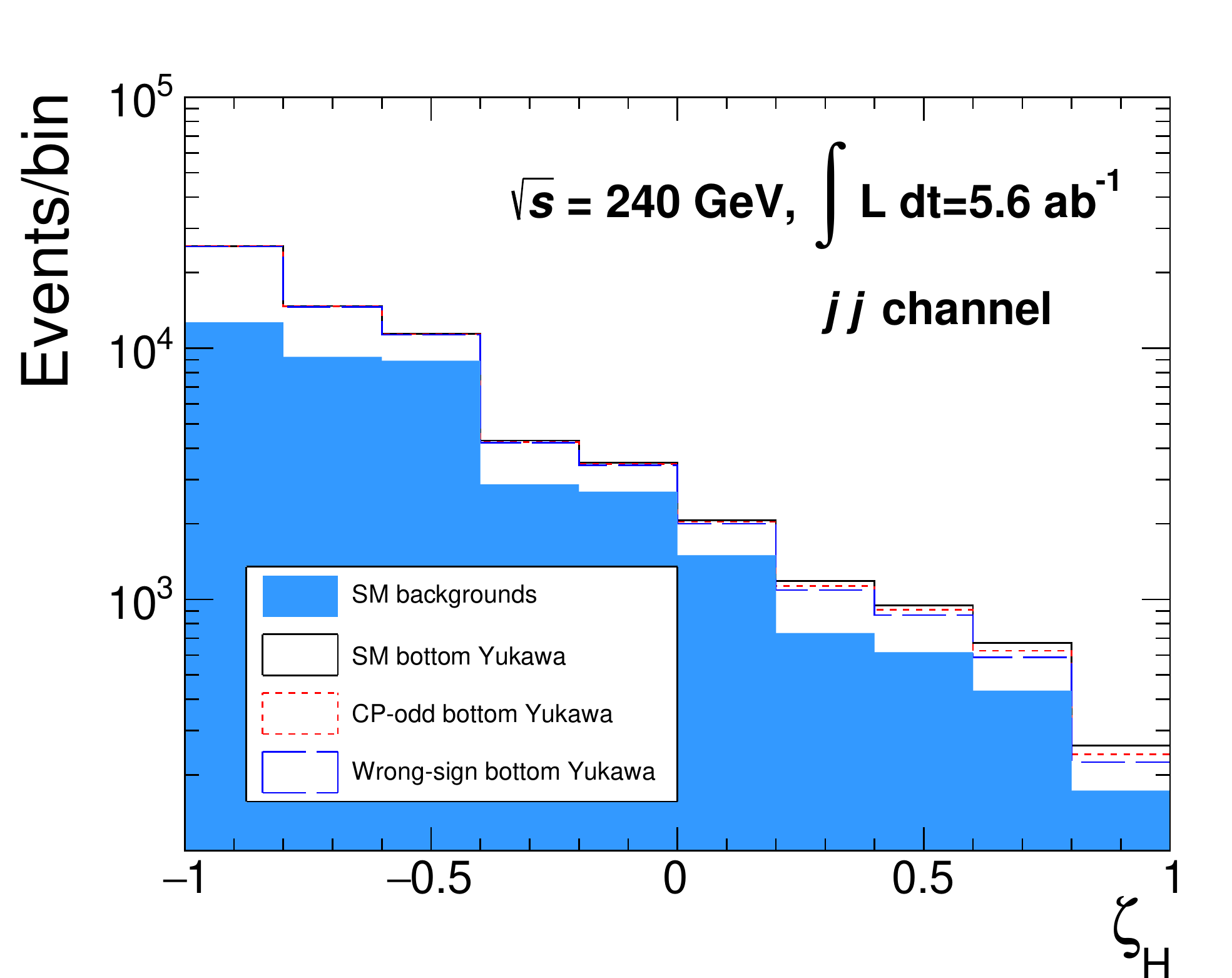}\\
\includegraphics[width=7.5cm]{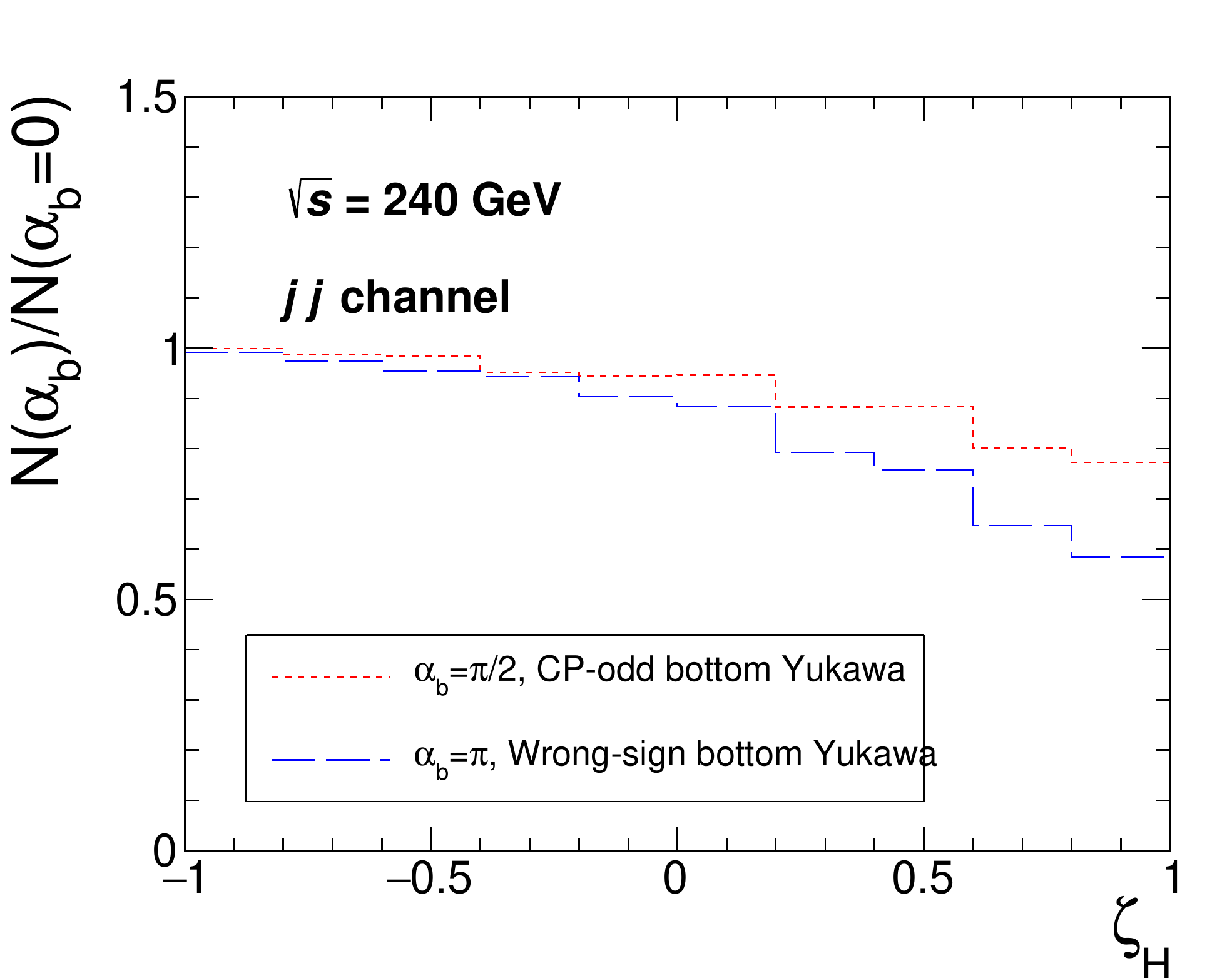}
\figcaption{\label{fig:3}The $\zeta_H$ distributions for the SM background, the SM bottom-quark Yukawa interaction ($\alpha_b=0$), bottom-quark Yukawa interaction with CP-odd scalar ($\alpha_b=\pi/2$), and the wrong-sign bottom-quark Yukawa interaction ($\alpha_b=\pi$) at 240 GeV Higgs factory with 5.6 ab$^{-1}$ integrated luminosity for hadronic decaying $Z$. {\it {Upper panel}}: The $\zeta_H$ distribution; {\it {Lower panel}}: The ratio of the event rates with respect to the SM case ($\alpha_b=0$). }  
\end{center}

\subsubsection{Data Analysis}
We define the binned likelihood function by 
\be
L(\mu,\alpha)\equiv\prod_{i=1}^{N_{\text{bin}}}\frac{\left[\mu s(\alpha)_i+b_i\right]^{n_i}}{n_i!}e^{-\mu s(\alpha)_i-b_i},
\label{eq:22}
\ee
where $\mu$ is the signal strength, $s(\alpha)_i$ is the numbers of the signal event in the $i$th bin under the hypothesis $\alpha_b=\alpha$, $b_i$ is the numbers of the SM background event in the $i$th bin, and $n_i$ is the number of total events observed in the $i$th bin. So under the assumption $\alpha_b=\alpha_0$, the logarithm of the ratio of the likelihood function will be
\bea
-2\Delta\log L&\equiv&-2\log\frac{L(\mu,\alpha)}{L(\mu_0,\alpha_0)}\nonumber\\
&=&-2\sum_{i=1}^{N_{\text{bin}}}\biggl\{\mu_0s(\alpha_0)_i-\mu s(\alpha)_i+[\mu_0 s(\alpha_0)_i+b_i]\nonumber\\
&&\times\log\left(\frac{\mu s(\alpha)_i+b_i}{\mu_0 s(\alpha_0)_i+b_i}\right)\biggr\}.
\label{eq:23}
\eea
With $-2\Delta\log L=q^2$ we may estimate the $q\sigma$ confidence level (C.L.) exclusion region under the SM hypothesis $\alpha_b=0$. We present the result in the complex plane of the complex parameter defined by $Y_b\equiv y_be^{i\alpha_b}/y_b^{\text{SM}}$. The result is shown in Fig. \ref{fig:4}.
\begin{center}
\includegraphics[width=7.5cm]{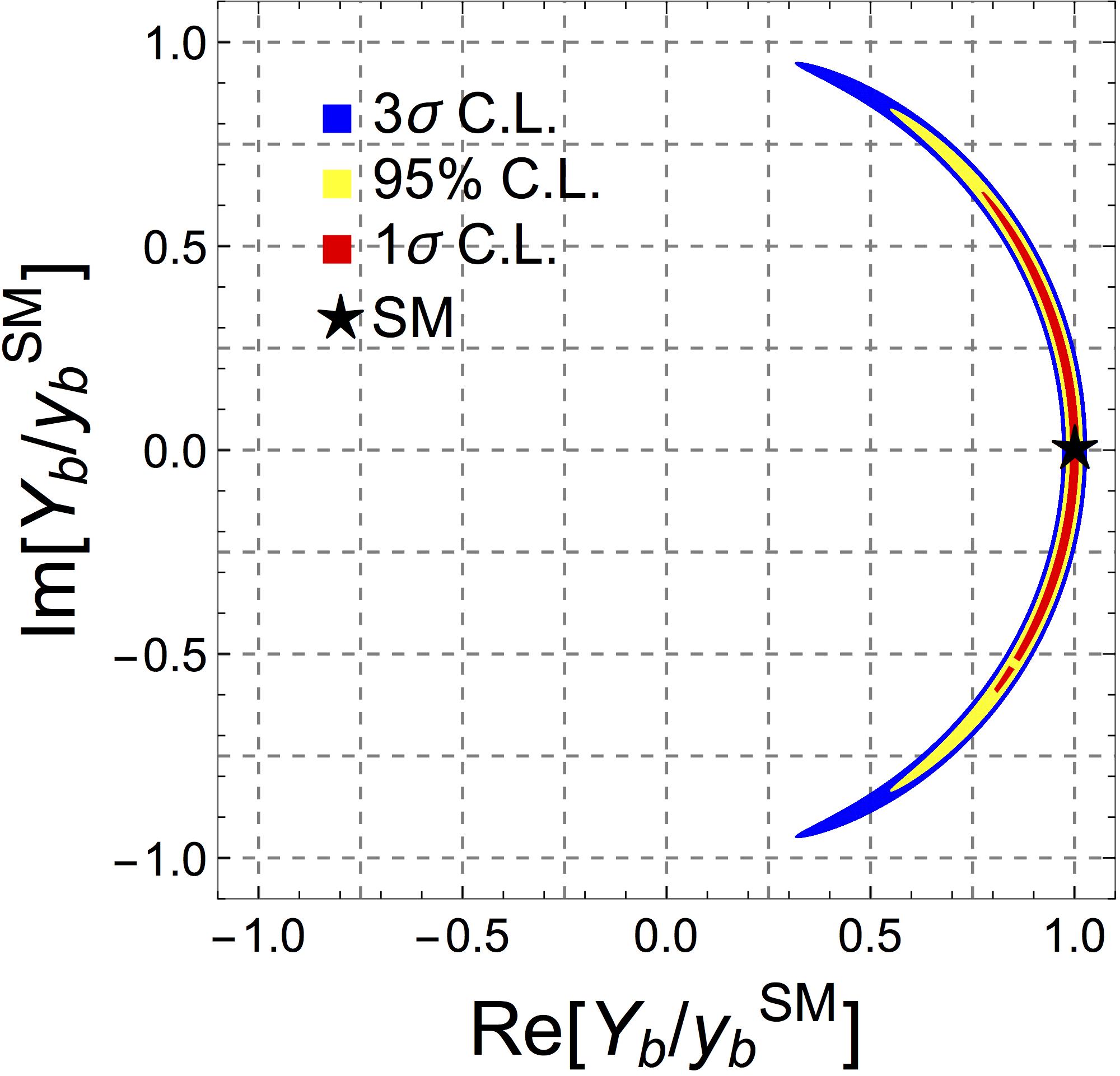}
\figcaption{\label{fig:4}The constraint to $Y_b$ at 240 GeV Higgs factory with 5.6 ab$^{-1}$ integrated luminosity after combining the leptonic and hadronic decaying $Z$ channels. }  
\end{center}

We may estimate the measurement uncertainty $\delta\alpha$ for arbitrary $\alpha_0$ with solving
\be
-2\log\frac{L(\hat \mu,\alpha_0+\delta\alpha)}{L(1,\alpha_0)}=1,
\label{eq:24}
\ee
where $\hat\mu$ is chosen by minimizing the quantity on left-hand side of Eq. (\ref{eq:24}). The result is shown in Fig. \ref{fig:5}. The larger uncertainty for $\alpha_b\to0$ and $\alpha_b\to\pi$ is due to the smaller derivative of the cosine function in these regions.
\begin{center}
\includegraphics[width=7.5cm]{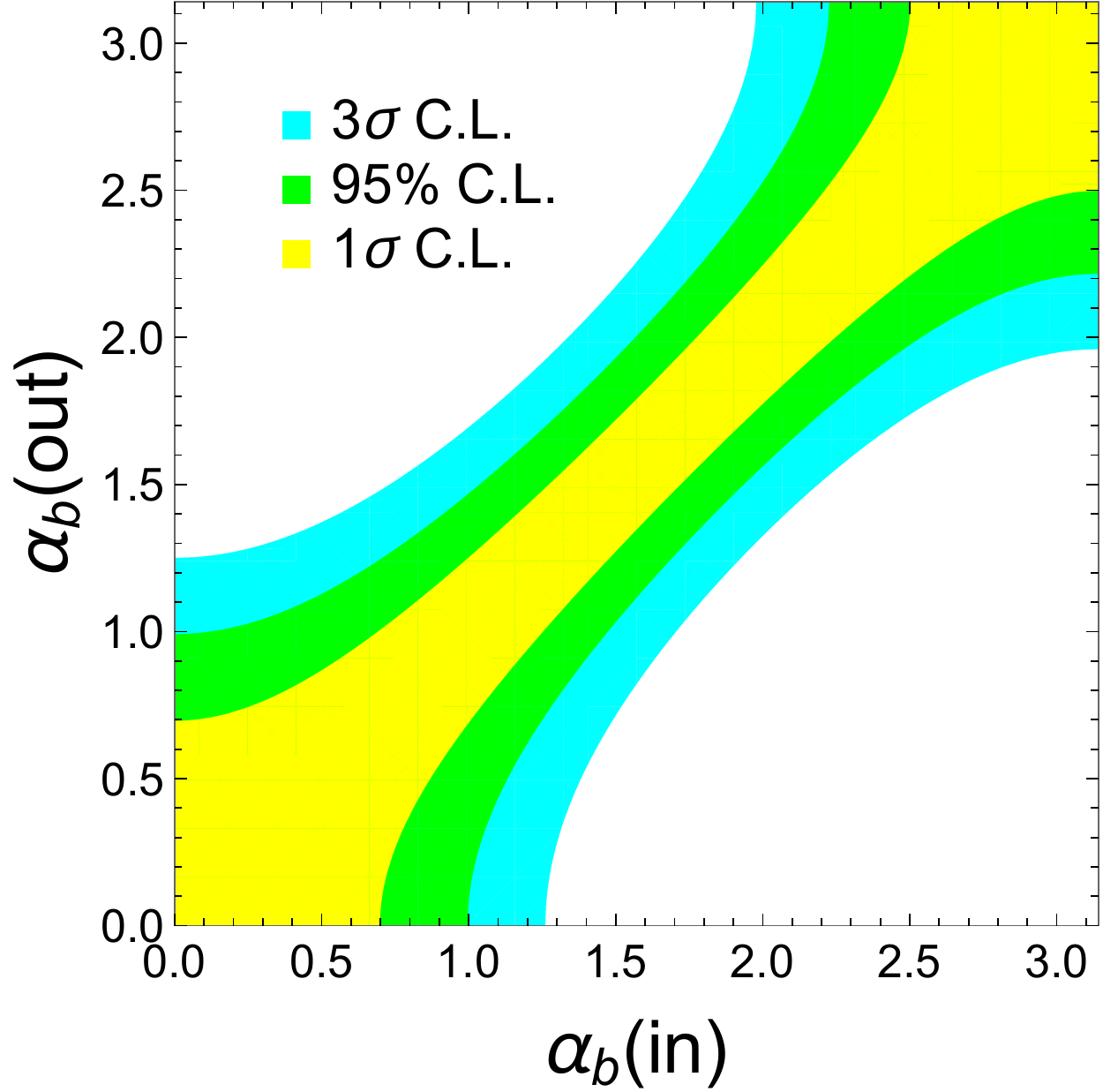}
\figcaption{\label{fig:5}The $\alpha_b$ measurement accuracy at 240 GeV Higgs factory with 5.6 ab$^{-1}$ integrated luminosity after combining the leptonic and hadronic decaying $Z$ channels. The $\alpha_b({\text{in}})$ is the real input of the phase angle, while $\alpha_b({\text{out}})$ is the measured value with uncertainty. }  
\end{center}

\subsection{The 365GeV $e^+e^-$ collider}
For the 365GeV $e^+e^-$ collider, we generate the events with the same method, choose the smearing parameters and the $k$-factor with the same value as the 240GeV Higgs factory, and use the same smearing formulas. The kinetic cuts are modified slightly. For the leptonic decaying $Z$ channel, the $\theta_{\ell^+\ell^-}$ cut is changed to $\theta_{\ell^+\ell^-}>60^\circ$. For the hadronic decaying $Z$ channel, the likelihood functions of the invariant mass distributions and recoil mass distributions are changed to 
\bea
L_Z(m)&=&P(m;91.1{\text{GeV}},5.58{\text{GeV}}),\label{eq:25}\\
L_h(m)&=&P(m;124.9{\text{GeV}},6.14{\text{GeV}}),\label{eq:26}\\
L_{rZ}(m)&=&P(m;131.88{\text{GeV}},23.84{\text{GeV}}),\label{eq:27}\\
L_{rh}(m)&=&P(m;102.6{\text{GeV}},30.27{\text{GeV}}),\label{eq:28}
\eea
and the recoil mass distributions do not help us a lot. Finally, we combine the result from 356GeV lepton collider with the result from the 240GeV Higgs factory shown before. The combined results are shown in Fig. \ref{fig:6} and Fig. \ref{fig:7}.
\begin{center}
\includegraphics[width=7.5cm]{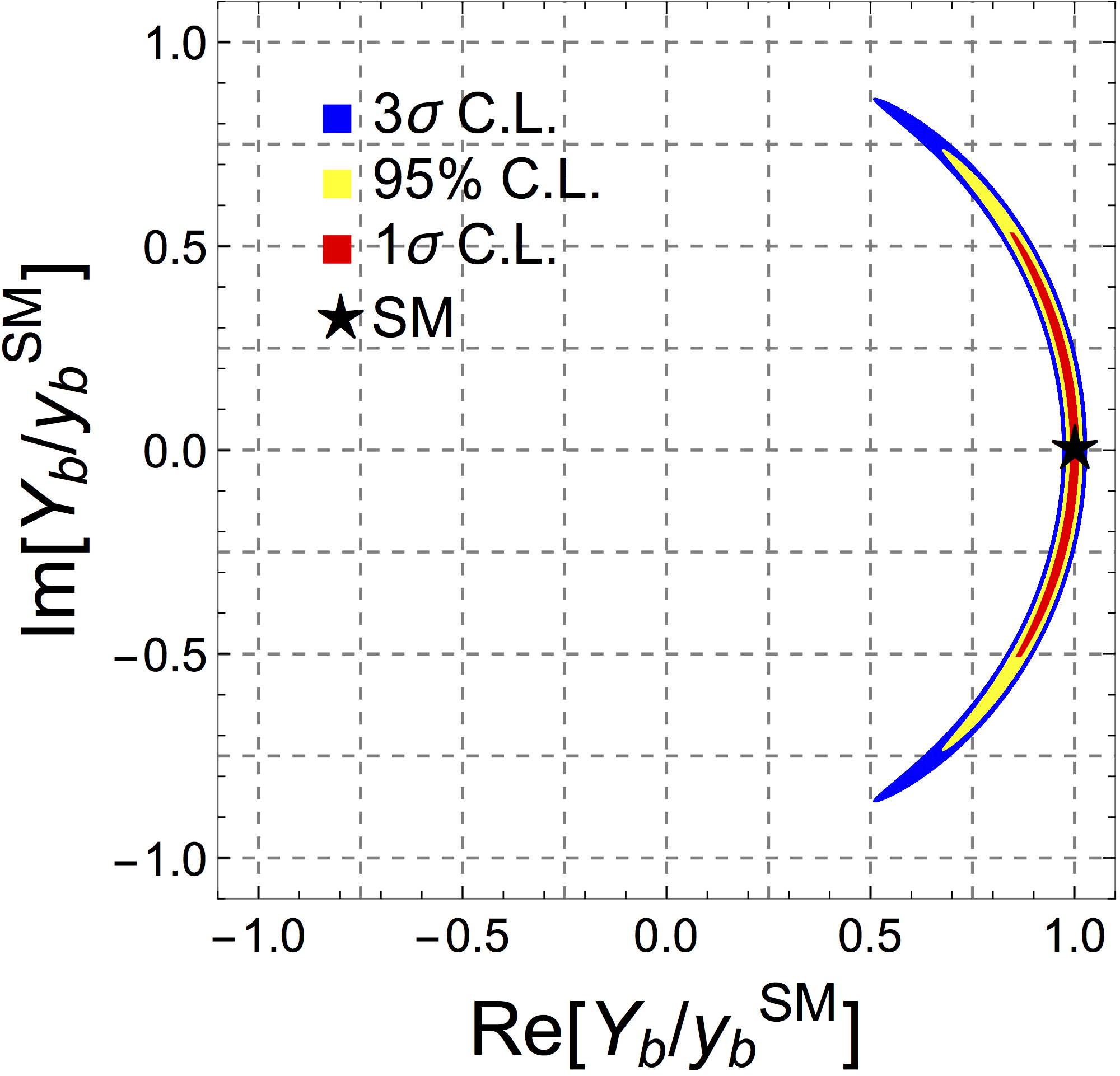}
\figcaption{\label{fig:6}The constraint to $Y_b$ at 240 GeV Higgs factory with 5.6 ab$^{-1}$ integrated luminosity combined with 365 GeV lepton collider with 1.5ab$^{-1}$ integrated luminosity after combining the leptonic and hadronic decaying $Z$ channels. }  
\end{center}
\begin{center}
\includegraphics[width=7.5cm]{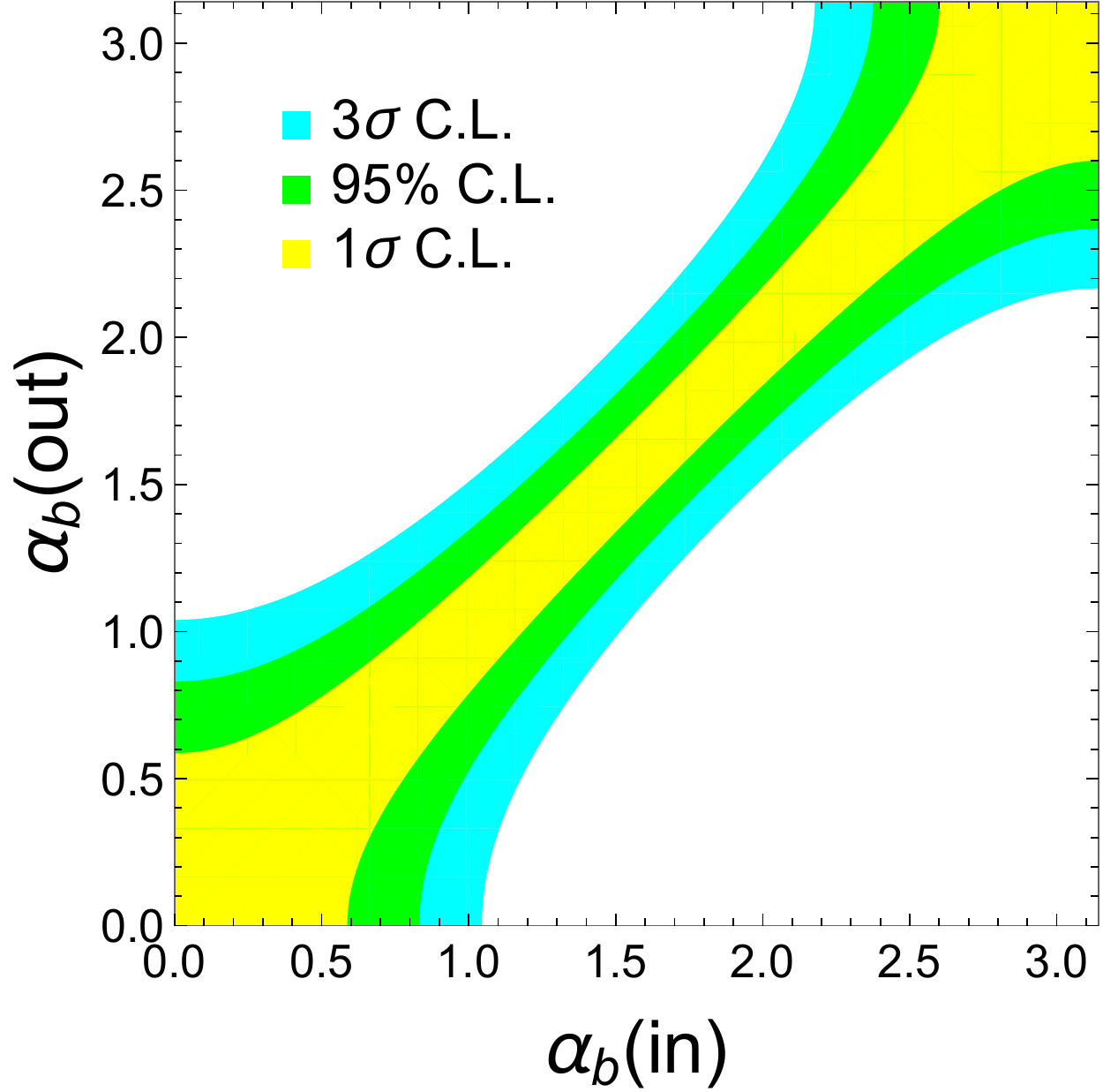}
\figcaption{\label{fig:7}The $\alpha_b$ measurement accuracy at 240 GeV Higgs factory with 5.6 ab$^{-1}$ integrated luminosity combined with 365 GeV lepton collider with 1.5ab$^{-1}$ integrated luminosity after combining the leptonic and hadronic decaying $Z$ channels. The $\alpha_b({\text{in}})$ is the real input of the phase angle, while $\alpha_b({\text{out}})$ is the measured value with uncertainty. }  
\end{center}

\section{Conclusion and Discussion}
In this work, we investigate the possibility of measuring the phase angle in the bottom-quark Yukawa interaction at future Higgs factory. We find that at 240 GeV Higgs factory with 5.6 ab$^{-1}$ integrated luminosity, the accuracy of this measurement could reach about $\delta(\cos\alpha_b)\sim \pm0.23$. If the Higgs factory will run at 365 GeV and accumulate 1.5ab$^{-1}$ integrated luminosity, the accuracy could increase to about $\delta(\cos\alpha_b)\sim \pm0.17$. This result, combining with the $hgg$ interaction measurement result from the LHC, can help us fix the phase angle in the bottom-quark Yukawa interaction with the 125 GeV SM-like Higgs boson discovered at the LHC.

In our simulation, we generate the Monte Carlo events with tree level amplitude. The infra-red (IR) divergency in the cross section is avoided by adding kinematic cuts. There are a lot of works on the higher order correction to the $h\to b\bar b$ decay channel since 1980s (for example, see \cite{Braaten:1980yq,Drees:1990dq,Djouadi:1994gf,Chetyrkin:1995pd,Larin:1995sq,Djouadi:1995gt,Butenschoen:2006ns,Anastasiou:2011qx,Gauld:2016kuu,Bernreuther:2018ynm,Primo:2018zby,Gao:2019mlt}). Some of them do include the interference effect with the $h\to gg$ channel. Because the phase space region which makes the dominant contribution to the measurement is the nearly collinear region of the two $b$-jets, a calculation including resummation effects in that region probably highly improve the accuracy of the theoretical prediction.  

The $b$-tagging efficiency used in this work is high. It is probably that the $b$-tagging efficiency at future Higgs factory could not reach the assumed value. There are some potential reasons which will decrease the $b$-tagging efficiency. For example, because the two $b$-jets are nearly collinear, it could be hard to tag both of them with high efficiency. Second, the $b$-jet in this process is not energetic enough so the mis-tagging rate of the charm-quark jet might be higher than our assumption. However, these will not be big problems. One may only require only one $b$-tagged jet in the signal events and accept a higher $c$-mis-tagging rate, because the simulation shows that these SM backgrounds are still small enough. When people try to analyze the data with hadronic decay $Z$ boson, these problems will be more subtle. A more realistic simulation is necessary in that case. Because the hadronic $Z$ decay branching ratio is much larger, those data might improve the result. Nevertheless, this topic is out of the scope of our work.

\acknowledgments {We thank Edmond L. Berger, Qing-Hong Cao, Lian-Tao Wang, Li Lin Yang, and Jiang-Hao Yu for helpful discussion. HZ would like to thank Shanghai Jiao-Tong University in Shanghai for hospitality. }

\end{multicols}

\vspace{-1mm}
\centerline{\rule{80mm}{0.1pt}}
\vspace{2mm}

\begin{multicols}{2}

\end{multicols}

\clearpage
\end{CJK*}
\end{document}